\documentclass[5p,times]{elsarticle}
\RequirePackage{etex}
\usepackage{amssymb}
\usepackage{amsthm}
\usepackage{makecell}
\usepackage{amsmath}
\usepackage{color,xcolor}
\usepackage{algorithm, algpseudocode}
\newtheorem{remark}{Remark}
\usepackage[utf8]{inputenc}
\usepackage{booktabs, caption, makecell}

\usepackage{threeparttable}
\usepackage{bbm}
\usepackage{tikz}
\usetikzlibrary{positioning}

\usepackage[markup=blue,deletedmarkup=sout,
addedmarkup=nf,final]{changes}
\usepackage{multirow}
\usepackage[figuresright]{rotating}
\usepackage{amsmath}
\newcommand{\R}{\mathbb{R}}
\usepackage{tablefootnote}
\newcommand{\Z}{\mathbb{Z}}
\DeclareMathOperator*{\minimize}{minimize}

\DeclareMathOperator*{\subt}{subject \ to}

\usepackage{tikz-network}

\usepackage{algorithm, algpseudocode}
\newtheorem*{AlgSel}{Algorithm selection}
\newtheorem*{AlgConf}{Algorithm configuration}

\usepackage[english]{babel}
\usepackage[utf8]{inputenc}
\usepackage{amsmath}
\usepackage{csquotes}

\begin{document}

\begin{frontmatter}

\title{Accelerating process control and optimization via machine learning: A review}

\author[1,2]{Ilias Mitrai\fnref{label2}}
\ead{imitrai@che.utexas.edu}
\author[1]{Prodromos Daoutidis\fnref{label2}} 
\fntext[label2]{Corresponding author}
\ead{daout001@umn.edu}
\address[1]{Department of Chemical Engineering and Materials Science, University of Minnesota, Minneapolis, MN 55455}
\address[2]{Currently with McKetta Department of Chemical Engineering, The University of Texas at Austin, Austin, TX 78712}

\begin{abstract} 
Process control and optimization have been widely used to solve decision-making problems in chemical engineering applications. However, identifying and tuning the best solution algorithm is challenging \added{and time-consuming. Machine learning tools can be used to automate these steps by learning the behavior of a numerical solver from data. In this paper,} we discuss recent advances in (i) the representation of decision-making problems for machine learning tasks, (ii) algorithm selection, and (iii) algorithm configuration for monolithic and decomposition-based algorithms. Finally, we discuss open problems related to the application of machine learning for accelerating process optimization and control.
\end{abstract}

\begin{keyword}
Process control \sep Process optimization \sep Machine Learning
\end{keyword}
\end{frontmatter}

\section{Introduction}
The design and operation of chemical processes depend on decisions spanning a wide range of scales, from the molecular up to the enterprise-wide, and constrained by multiple physical and chemical phenomena \cite{hanselman2016mathematical, grossmann2012advances, daoutidis2018integrating, pistikopoulos2021process}. Process control and optimization methods provide a systematic framework to identify the best possible decisions in designing and operating a process, subject to constraints that emerge from physics or design and operational considerations. Over the last few decades, there have been significant advances in both theory and algorithm development regarding the control of nonlinear and constrained process systems \cite{daoutidis2023future, christofides2013distributed, ellis2014tutorial, mcallister2022advances, mesbah2016stochastic, shin2019reinforcement}, as well as the solution of broad classes of optimization problems \cite{wachter2006implementation, tawarmalani2005polyhedral,boukouvala2016global, biegler2024multi, grossmann2016recent}. 

Despite these advances, control and optimization problems that challenge the computational performance of state-of-the-art algorithms continue to emerge. Some examples of application domains where such problems occur include the real-time operation of chemical processes interacting with renewable energy resources, the decarbonization of the energy sector, and the design of resilient, sustainable and circular supply chain networks \cite{national2022new}. The scale and complexity in these systems and the multiple spatial and temporal scales that are often present make the solution of the corresponding control and optimization problems challenging. Different approaches have been followed to improve the tractability of such problems. For example, one can potentially reduce the computational complexity by reformulating the problem \cite{raghunathan2003mathematical, liberti2006exact}. However, finding a suitable exact reformulation is generally not possible. Data-driven approaches namely surrogate and hybrid modeling, have also been developed to learn a surrogate model with lower computational complexity \cite{cozad2014learning, bhosekar2018advances, bradley2022perspectives, sansana2021recent, misener2023formulating}. Although this approach has received significant attention, the solution returned is (inherently) approximate. 

An alternative approach is to accelerate the solution process itself by 1) selecting a solution strategy (algorithm selection) and 2) tuning it (algorithm configuration) such that a desired performance function like solution time is minimized. The acceleration is usually achieved by exploiting some underlying property of the decision-making problem. An example is the case of structured decision-making problems, where the structure can be used as the basis of decomposition-based optimization algorithms, which are usually faster than monolithic algorithms for large-scale problems \cite{conejo2006decomposition}. Although this approach does not compromise solution quality, selecting and tuning a solution algorithm is nontrivial. Current state-of-the-art algorithms or solvers, especially commercial ones, are complex \deleted{black-box} systems with many algorithmic steps, each one potentially having a set of hyperparameters. Furthermore, the quantitative effect of the problem formulation on the performance of an algorithm, such as solution time, is not known a-priory, i.e., the selection and tuning of the algorithm are black-box optimization problems \added{since the solution time or quality (for local solvers) can not be determined apriori}. 

To this end, Machine Learning (ML) can be used to learn the behavior of an algorithm for a class of decision-making problems from data. ML has been widely used in chemical engineering for modeling chemical and physical systems and developing data-driven optimization and control algorithms \cite{cozad2014learning, bhosekar2018advances, bradley2022perspectives, sansana2021recent, daoutidis2023machine, schweidtmann2021machine, tang2022data}. Usually, data is used to learn a system's chemical, physical, or control-relevant properties. In the context of algorithm selection and configuration, the data are used to learn the effect of the problem formulation on the computational performance of an algorithm. 

The application of ML for accelerating an algorithm has recently received significant attention in the operations research and computer science communities and has shown the potential for significant computational savings  \cite{bengio2021machine}. This approach has received less attention in the chemical engineering literature, where the emphasis has been on improving the problem formulation and developing new optimization algorithms with well-characterized optimality properties. ML has been mainly used to analyze the solution time for production scheduling optimization problems \cite{kim2022supervised} and accelerate decomposition-based algorithms for the solution of mixed integer model predictive control \cite{mitrai2024initialize, mitrai2024computationally}, supply chain optimization \cite{triantafyllou2024deep}, and capacity expansion problems \cite{allen2023improvements}. \added{Decision-making problems that arise in chemical engineering have certain features, such as nonlinearity in the form of bilinear terms (flowrate multiplied by concentration) or exponentials with continuous variables $(e^{-\frac{E}{RT}}c)$ and certain structure in the constraints, such as tri-diagonal structure which arises in model predictive control applications.} {\it We posit that developing ML-based methods for accelerating general-purpose solvers as well as decomposition -based soltuion algorithms is a fitting approach to improve the tractability of complex decision-making problems in chemical engineering}.

In this paper, we aim to review the algorithm selection and configuration problems, review recent advances in using ML to accelerate the solution of decision-making problems and discuss open problems and future directions for applying this approach to chemical engineering problems. In Section~\ref{section: alg sel and conf probs}, we formally introduce the algorithm selection and configuration problems. In \added{S}{\deleted{s}ection~\ref{sec: representation of optimization problems}, the representation of an optimization problem in a format that can be used as input to standard ML models is discussed. In Section~\added{\ref{sec: learn to select}-\ref{tune decomposition algorithms}}\deleted{\ref{sec: learn to conf}, \ref{sec: learn to select}}, we present the application of ML for selecting and tuning an algorithm, and finally, in Section~\ref{sec: conclusions}, we discuss open problems and opportunities related to the acceleration of numerical algorithms using machine learning.

\section{The algorithm configuration and selection problems} \label{section: alg sel and conf probs}
\subsection{The algorithm selection problem} \label{sec: alg sel}
Consider a general decision-making problem
\begin{equation} \label{full model}
	\begin{aligned}
			P(p) := \minimize_{x} \ \ & f(x;p) \\
			\subt \ \ & g_i(x;p) \leq 0 \ \ \forall i=1,...,m_{in}\\
			& h_j(x;p) = 0 \ \ \forall j=1,...,m_{eq}\\
			& x \in \R^{N_x^c}\times \Z^{N_x^d},
		\end{aligned}   
\end{equation}
where $N_x^c + N_x^d = N $, $ m_{in}+m_{eq}=M$, and $p$ are the parameters of the problem \added{(which can be both continuous and integer)}, $x$ are the decision variables and the objective $f$ as well as the constraints $g_{i},h_{j}$ can be convex \added{(linear or nonlinear)} or nonconvex. The first question that arises during the solution of a decision-making problem is which algorithm to select for the solution of the problem. In general, finding or developing an algorithm that performs well for any decision-making problem is not possible \cite{markot2011comparison, smith2012measuring, hutter2014algorithm}. Thus, for a given problem, one must find the \deleted{best} \added{most suitable} algorithm or solution strategy. This is formally known as the algorithm selection problem and is stated as follows \cite{rice1976algorithm}:
\begin{AlgSel}
    \normalfont Given an optimization problem $P(p)$ and a set of algorithms $\mathcal{A} = \{a_1,..., a_{|\mathcal{A}|}\}$, determine which algorithm $\alpha^*$ should be used to solve the problem such that a desired performance function $m: \mathcal{P} \times \mathcal{A} \rightarrow \mathcal{M}$ is optimized. 
\end{AlgSel}
\noindent The performance function $m$ is a metric used to compare two algorithms. Typical performance functions can be the computational time or the solution quality for a given computational budget. The choice of the performance function depends on the application. For example, solution time might be more important for an online application, whereas solution quality and feasibility might be better for a design or safety-critical application. 

Given a decision-making problem, the set of available algorithms, and a performance function, algorithm selection can be posed as an optimization problem as follows
\begin{equation} \label{eq: alg sel prob def}
    \alpha^{*} \in \arg \min_{\alpha \in \mathcal{A}} \ m(P(p), \alpha).
\end{equation}
This problem is also known as per-instance algorithm selection since it considers only a specific decision-making problem. However, it can be easily extended to identify the best algorithm for a class of decision-making problems \cite{kerschke2019automated}.

The algorithm selection problem is a black-box optimization problem since the performance function $m$ is not known explicitly, and evaluating an algorithm for a given problem can require significant computational resources. The standard approach to solving this problem relies on ML, where data are used to approximate the performance function, and the best algorithm is selected based on the predictions of the learned model.

\subsection{Algorithm configuration} \label{sec: alg conf}
Once an algorithm is selected, the next step is tuning of the algorithm. Let's consider the case where an algorithm $\alpha$ with parameters $\pi_{\alpha}$ is available to solve a decision-making problem $P(p)$ (Eq.~\ref{full model}). We will refer to the parameters of the algorithm $\pi_{\alpha}$ as hyperparameters in order to distinguish them from the parameters of the decision-making problem $p$. The values of the hyperparameters $\pi_{\alpha}$, also known as tuning or configuration, have a significant effect on the computational performance of the algorithm. Usually, these hyperparameters are selected by considering the average performance of the algorithm over a set of instances. However, one can exploit specific features of a  problem and find a tuning that is optimal for the specific instance. This problem is formally known as the per-instance algorithm configuration problem and is stated as follows \cite{eggensperger2019pitfalls, schede2022survey}:
\begin{AlgConf}
    \normalfont Given a decision-making problem $P(p)$, and an algorithm $\alpha$ with hyperparameters $\pi_{\alpha} \in \Pi_{\alpha}$ find hyperparameters $\pi^{*}_{\alpha}$ such that a performance function $m_{conf}^{\alpha} : \mathcal{P} \times \Pi_{\alpha} \mapsto \mathcal{M}$ is optimized.
\end{AlgConf}
The algorithm configuration problem has three components. The first is the decision-making problem $P(p)$, which is given. The second is the space of possible configurations $\Pi_{\alpha}$, which is algorithm dependent. For example, in gradient descent algorithms, a common hyperparameter is the step size (or learning rate), which is a positive number, i.e., $\Pi_{\alpha} = \mathbb{R}_{+}$. The last component is the performance function $m_{conf}^{\alpha}$, a metric used to compare two configurations of the algorithm $\alpha$ for a given problem. Similar to the algorithm selection problem, based on the application considered, different performance functions can be used  such as solution time or solution quality. These components lead to the following formulation of the algorithm configuration problem
\begin{equation} \label{eq: alg conf prob}
\pi^{*}_{\alpha} \in \arg \min_{\pi_{\alpha} \in \Pi_{\alpha}} \ \ m_{conf}^{\alpha} (P, \pi_{\alpha}).
\end{equation}
The solution to the algorithm configuration problem is challenging. First, the performance function $m_{conf}^{\alpha}$ is not known explicitly, i.e., algorithm configuration is a black-box optimization problem. Also, evaluating the performance of a configuration for a given problem can be computationally expensive for large-scale decision-making problems. Finally, the search space of possible algorithm configurations can be very large. 

The first approach to solving the algorithm configuration problem is to rely on sampling-based black-box optimization algorithms. Although this approach has been extensively used in the literature \cite{liu2019tuning, chen2011random, hutter2009paramils, hutter2011sequential, hutter2010automated}, it can be slow for online applications, where given a decision-making problem, one must quickly find the best configuration of the algorithm and implement it. In such cases, ML can be used to learn (or approximate) offline either the performance function $m_{conf}^{\alpha}$ using a surrogate $\hat{m}_{conf}^{\alpha}$ or the solution of the algorithm configuration problem itself. Once these models are learned, then they are used online to find the best configuration.

\subsection{Relation between algorithm selection and configuration}

The algorithm selection and configuration problems share several characteristics. First, algorithm configuration can be considered as a special case of algorithm selection. Specifically, each configuration of an algorithm can be considered as a different algorithm, and thus, identifying the best possible configuration is equivalent to selecting the best algorithm. \added{This can be considered as a simultaneous algorithm selection and configuration approach since one must consider simultaneously all the possible combinations of algorithms and configurations.} In general, algorithm configuration is usually more challenging than algorithm selection since the search space is much larger. The algorithm selection problem has one degree of freedom,  the algorithm to be used, and the number of available algorithms is usually small. However, in the algorithm configuration problem, the degrees of freedom are equal to the number of hyperparameters, and the possible number of configurations can be very large.

Both problems can be solved either via black-box optimization or ML-based approaches. In general, black-box methods have been used for offline applications where a solver is tuned to perform well on average for a given set of instances. Black-box optimization methods require a function evaluation, i.e., computing the performance function $m$ for a given problem. In the context of algorithm selection and configuration, this translates to solving the decision-making problem $P(p)$ to optimality and obtaining the value of the performance function. This approach can not be applied in an online setting where one has to identify the best algorithm or tuning without solving the problem. In such cases, one must learn a surrogate model from data offline and then use it for inference online.

Finally, the tasks of selecting and tuning an algorithm, as presented in Sections~\ref{sec: alg sel} and \ref{sec: alg conf}, can be considered static problems since they are solved only once. In general, algorithm selection and configuration can be performed multiple times when solving a decision-making problem, leading to dynamic algorithm selection and configuration problems. \added{Consider, for example, branch and bound-based algorithms where different solvers can be used at different nodes in the tree \cite{markot2011comparison} or even different solver tuning.} This difference \added{(static or dynamic)} motivates the adoption of different solution strategies (see Fig.~\ref{fig:alg sel and conf framework}). The static case is a one-step decision-making problem since, given an optimization problem (Eq.~\ref{full model}), the ML model is used to identify the best algorithm or tuning. On the contrary, the dynamic case requires constant interaction between an ML model and the algorithm. Given the decision-making problem (Eq.~\ref{full model}) and the state of the solution process, the ML model determines the best configuration for the algorithm; this is a multi-step decision-making problem. This difference motivates the application of supervised and reinforcement learning for algorithm selection and configuration as presented in the next sections.  

\begin{figure*}
    \centering
    \includegraphics[scale=1]{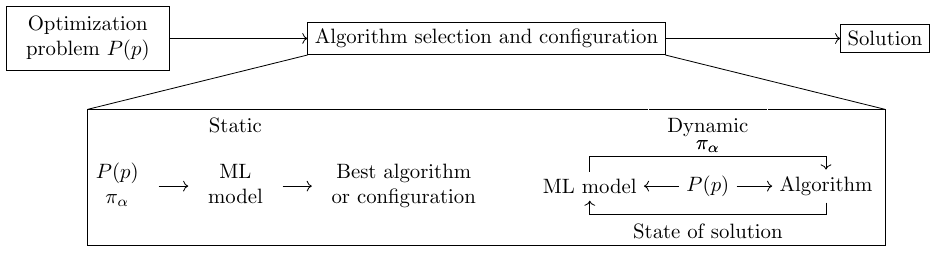}
    \caption{High level overview of ML-based solution approaches for algorithm selection and configuration}
    \label{fig:alg sel and conf framework}
\end{figure*}

\section{Decision-making problems as inputs to ML models} \label{sec: representation of optimization problems}
Let's consider the case where an ML model $\hat{m}$, such as a feedforward neural network, is used to predict the solution time $t$ of a given algorithm $a$ for an instance $P(p)$, i.e., $t=\hat{m}(P(p), \alpha)$. A major limitation in developing such a model is that the optimization problem in Eq.~\ref{full model} can not be used as the input to standard ML models, such as a neural network, decision tree, random forest, etc. A decision-making problem can not be considered as a tabular or euclidian data point since it has variables, constraints, and an objective (or multiple ones in the case of multi-objective problems). Also, the number of variables and constraints can vary for different problems or instances. Therefore, a transformation step is necessary to represent a decision-making problem in a form that can be used as the input to an ML model. This representation should 1) capture essential information about the problem, 2) be amenable to use for different problem sizes, i.e., varying number of variables and constraints, 3) not be affected by the ordering of the variables and constraints, and 4) be computed/constructed efficiently. 

\subsection{Vectorial feature representation} \label{sec: vectorial representation}
The standard approach to achieve the above requirements is to extract a set of easily computable features $\nu(P(p)) \in \mathbb{R}^{N_{features}}$ from the problem formulation and use them as inputs to an ML model (see Fig.~\ref{fig:feature representation}). We will refer to this as the vectorial feature representation. Examples of these features include the number of continuous and discrete variables, the number of constraints, the number of nonconvex terms, the convexity of the objective, etc. We refer the reader to \cite{ smith2012measuring, hutter2014algorithm, schede2022survey} for an extended list. Although this approach has been extensively used to predict the solution time of mixed integer optimization problems, it has two limitations. First, significant effort and domain knowledge are required to identify the most informative features for a given class of problems. Secondly, the vectorial representation does not account for the exact interaction pattern among the variables and constraints.
\begin{figure}[h]
     \centering
     \includegraphics[scale=0.8]{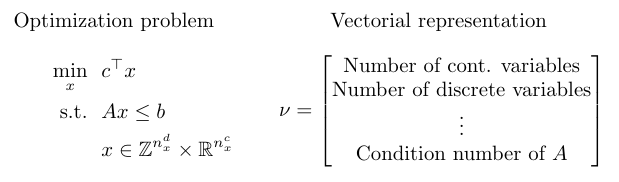}
     \caption{Vectorial feature representation of an optimization problem}
     \label{fig:feature representation}
 \end{figure}
 
\subsection{Graph Representation} \label{sec: graph representation}
An alternative approach to represent a decision-making problem is a graph that can capture the interaction pattern between the variables and constraints. A graph $\mathcal{G}$ is a mathematical object that captures the interaction between a set of objects called nodes or vertices. We define node $i$ as $v_i$ and $V = \{v_{i}\}_{i=1}^{N}$ the set of nodes. The interaction pattern is captured via the edges $E = \{e_{ij}\}_{i \in V, j \in V}$ where $e_{ij}=1$ if node $i$ is connected with node $j$. A graph can also be represented by the adjacency matrix $A \in \mathbb{R}^{N \times N}$ where $A_{ij}=1$ if an edge exists between node $i$ and $j$, i.e., if $e_{ij}=1$. 

Three graphs can be used to represent a decision-making problem \cite{allman2019decode}. The first and most generic one is the bipartite variable-constraint graph $\mathcal{G}_{b} (V_{n}, V_{m}, E)$ $(|V_{n}|=n, |V_{m}|=m)$ with adjacency matrix $A_b$. This graph has two sets of nodes, one representing the constraints $V_{m}$ and the other the variables $V_{n}$. The edges $E$ capture the presence of a variable in a constraint. The second type of graph is a constraint graph $\mathcal{G}_{c} (V_{m}, E_{m})$, where the nodes $V_{m}$ are the constraints of the problem and the edges $E_{m}$ \added{represent the variables that couple two constraints.} \deleted{represent whether two constraints are coupled by involving one or more common variables.} In this case, an edge between two nodes $i$ and $j$ can have a weight $w_{ij} \in \mathbb{Z}_{+}$, which denotes the number of variables that couple two constraints. The third type is the variable graph $\mathcal{G}_n (V_n, E_n)$ where the nodes $V_n$ are the variables of the problem, the edges $E_m$ \added{represent whether two variables are coupled by appearing together in one or more constraints} \deleted{represent the constraints that couple two variables}, and each edge has a weight that denotes the number of constraints that couple two variables. An example of a variable graph is presented in Fig.~\ref{fig:graph representation}.

The graph representation captures the structural coupling between the constraint and the variables, i.e., the presence or not of a variable in a constraint, as reflected in the adjacency matrix, as well as the strength of interaction captured via the edge weights. Such representations have been used extensively for developing control architectures, as well as implementing decomposition-based optimization \added{and control} algorithms \added{\cite{allman2019decode, tang2018optimal, jogwar2017community, moharir2017graph, michelena1997hypergraph, wang2013computational, ferris1998partitioning, aykanat2004permuting, jalving2022graph, bergner2015automatic, khaniyev2018structure, del2016automated, shin2021graph}}. 

Under this representation, a decision-making problem, and in general a system of equations, is represented by a graph $\mathcal{G}$ with adjacency matrix $A$. Note that the graph and the adjacency matrices depend on the decision-making problem $P(p)$, i.e., $G(P(p))$ and $A(P(p))$.
\begin{figure}[h]
     \centering
     \includegraphics[trim = 10 0 0 0, clip,scale =0.8]{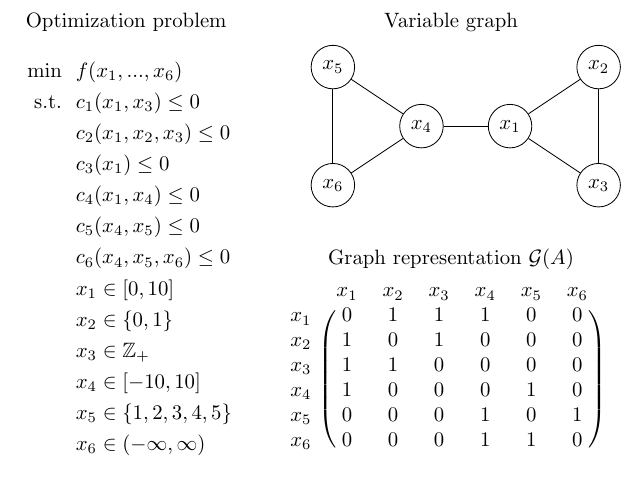}
     \caption{Graph representation of an optimization problem}
     \label{fig:graph representation}
 \end{figure}

\subsection{Graph representation with nodal and edge features} \label{sec: graph with features representation}
Although the graph representation captures the structure of the problem, it does not account for the domain of the variables and the functional form in which they appear in the constraints. To achieve this, a set of features can be associated with each node and edge in the graph. For example, in the bipartite graph representation, a set of features $\phi_{v}^{i}$ can be used for each variable $i$, $\phi_{c}^{j}$ for each constraint $j$, and $\phi_{e}^{ij}$ for an edge between variable $i$ and constraint $j$. Concatenation of these features form the feature matrices $F_{v}$, $F_{c}$, $F_{e}$, and a decision-making problem (Eq.~\ref{full model}) can be represented by four matrices, the adjacency matrix $A$, the variable feature matric $F_{v}$, the constraint feature matrix $F_{c}$, and the edge feature matrix $F_{e}$ (see Fig.~\ref{fig:graph representation with features} for an example).

This representation has been extensively used for Mixed Integer Linear Programming problems \cite{gasse2019exact, ding2020accelerating, gupta2020hybrid, gupta2022lookback, li2022learning, liu2022learning, nair2020solving, paulus2022learning}. Some examples of features include the domain of the variables for the variable nodes, the type of constraint (equality or inequality) for the constraint nodes, and the coefficient of a variable in an edge for the edges. The ability of this representation to distinguish between different optimization problems has been proven rigorously for specific classes of LP and MILP problems and for specific tasks such as predicting optimal solution and feasibility \cite{chen2022representinglp, chen2022representingmilp}.

 \begin{figure}
     \centering
     \includegraphics[trim= 10 0 0 0, clip,scale=0.65]{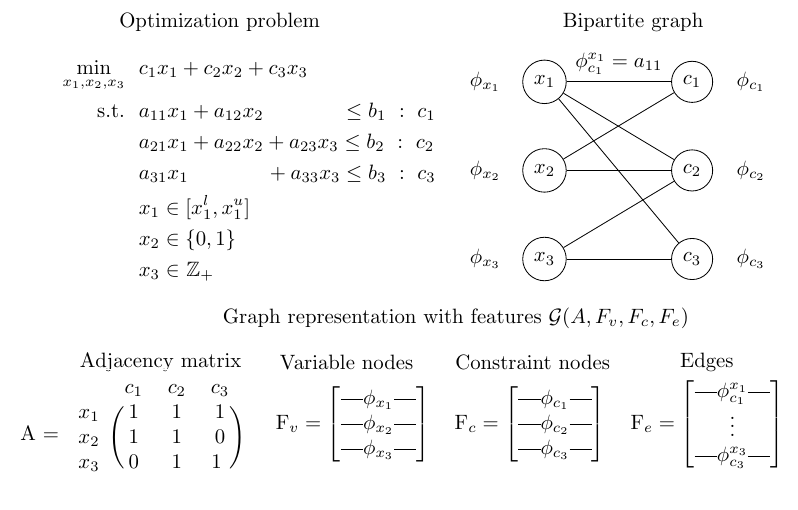}
     \caption{Graph representation with features of a mixed integer linear optimization problem}
     \label{fig:graph representation with features}
 \end{figure}

\begin{remark}
    \normalfont The representations presented in this section can be used as inputs to a surrogate model that predicts the computational performance of an algorithm. To this end, the following question arises: \textit{Which representation should be used?} The chosen representation should be able to represent the key characteristics of a problem that affect the computational performance of a solver. Furthermore, the selection of the representation will determine the class of ML models that can be used. The vectorial representation can be used with interpretable \deleted{modes} \added{models}, such as decision trees and linear regression, as well as noninterpretable models, such as neural networks, random forests, gaussian processes, etc. The graph representation requires geometric deep learning models \cite{bronstein2017geometric, bronstein2021geometric}, such as graph neural networks, which are not inherently interpretable. This selection affects our ability to understand the computational performance of a solver (see Section \ref{subs: new insights} for a detailed discussion on this). 
\end{remark}

\section{Learning to select a solution strategy} \label{sec: learn to select}

Given the aforementioned representations, first, we focus on the application of ML techniques for algorithm selection. One approach relies on regression to predict the value of the performance function for a given problem and then select the best algorithm. For each available algorithm $\alpha$, data are generated to approximate the performance function $m_{\alpha}$ with a surrogate $\hat{m}_{\alpha}$ where the input is a representation of the decision-making problem and the label is the value of the performance function of algorithm $\alpha$. \added{In this data generation process, the tuning of each algorithm $\alpha$ can either be the default one or the best possible one for the given instance.} This approach has mainly exploited the vectorial feature representation of a decision-making problem \cite{hutter2014algorithm, leyton2009empirical} to predict the solution time of algorithms using neural networks \cite{smith2011discovering, eggensperger2017neural}, decision trees \cite{bartz2004tuning}, gaussian processes \cite{smith2011discovering}, and sparse polynomial regression \cite{huang2010predicting}. Some applications include determining if dynamic programming or branch and search should be used for solving a knapsack problem \cite{hall2007performance} and selecting a heuristic for constraint programming \cite{allen1996selecting}.

An alternative approach is to approximate the solution of the algorithm selection problem itself, i.e., approximate the mapping $\mathcal{C}$ between the decision-making problem $P(p)$ and the best algorithm with a surrogate one $\hat{\mathcal{C}}$, i.e., $\alpha^{*} = \hat{\mathcal{C}}(P(p))$. In this approach, the output of the approximate map $\hat{\mathcal{C}}$ is one of the available algorithms. Thus, the algorithm selection problem can be posed as a multi-class classification problem where a classifier will predict the solver that has the highest probability of being the solution to the algorithm selection problem. This approach has been used to determine the best solution strategy for traveling salesman problems \cite{pihera2014application}, select local nonlinear solver during branch and bound for mixed integer nonlinear optimization problems \cite{markot2011comparison}, determine whether Dantzig-Wolfe decomposition should be used for the solution of mixed integer linear optimization problems \cite{kruber2017learning}, determine whether a convex mixed integer nonlinear optimization should be solved with branch and bound or the outer approximation algorithm \cite{mitrai2024taking}, and determine whether a mixed integer quadratic optimization problem should be linearized or not \cite{bonami2022classifier}.

Yet another solution approach is based on case-based reasoning, an artificial intelligence approach where a task is solved based on the solution of other similar tasks \cite{kolodner1992introduction}. In the context of algorithm selection, for a given problem, an algorithm $\alpha$ is selected based on its performance in similar instances. Case-based reasoning has been used to select whether a constraint programming or mixed integer programming approach should be used to solve a bid evaluation problem in combinatorial auctions using as features some properties of the graph representation of the problem, such as graph density, node degree, etc. \cite{guerri2004learning}. 

\section{Learning to configure an algorithm} \label{sec: learn to conf}
The problem of learning to configure an algorithm has received significant attention from the operations research, computer science, and ML/AI communities. We hereby focus only on the acceleration of optimization algorithms for the solution of linear, mixed integer linear, and mixed integer nonlinear optimization problems. A decision-making problem can be solved either monolithically, where an algorithm considers all the variables simultaneously, or using a decomposition-based algorithm, where the problem is decomposed into a number of subproblems that are solved iteratively. Given the different nature of monolithic and decomposition-based algorithms, different algorithm configuration tasks arise. Thus, we consider the configuration of these algorithms separately.

\subsection{Configuring monolithic solvers}
\subsubsection{Initialization} \label{subsub:initialization}
Initialization of an optimization algorithm is not usually considered a hyperparameter, yet it can have a significant effect on its computational performance. Usually, intuition and heuristics are used to identify a good feasible solution. However, the development of such initialization approaches is time consuming. 

ML has been used to predict the optimal solution of a class of decision-making problems and use the prediction either as an initial guess or to fix some of the variables of the problem. This approach relies on input-output data $\mathcal{D} = \{ P(p_{i}), x^{*}_{i}\}_{i=1}^{N_{data}}$, where the features are some representation of the decision making problem, as discussed in Section~\ref{sec: representation of optimization problems}, and the label is the optimal solution $x^{*}_{i}$ or part thereof. Given such data sets, supervised learning is used to train regression and classification models. Usually, regression is used for predicting the values of continuous variables, whereas classification is used for integer variables. This approach has been extensively used to accelerate the solution of decision making problems which are solved repeatedly online. Typical examples include \deleted{mixed integer} model predictive control (MPC), where ML models predict \added{the control action \cite{vaupel2020accelerating, kumar2021industrial}}, the values of the integer variables \cite{masti2020learning, zhu2020fast, cauligi2022prism, russo2023learning} (for mixed integer MPC), active constraints \cite{klauvco2019machine, cauligi2021coco,misra2022learning, bertsimas2022online}, optimal power flow problems \cite{park2023self}, and facility location problems \cite{triantafyllou2024deep}.

An alternative approach is to approximate the iterative nature of optimization algorithms via ML models, i.e., emulate the evolution of the variables' values during the solution process. This approach has been used to emulate interior point solvers for predicting the solution of optimal power flow problems \cite{baker2022emulating} using the feature representation and general linear optimization problems \cite{qian2024exploring} using the graph representation of the problem with features. These initialization approaches are based on the assumption that an initial guess close to the optimal solution will reduce the computational time. The main limitation of these initialization approaches is that the prediction is not necessarily feasible. Therefore, a feasibility restoration step is required to construct a feasible solution \cite{kotary2021end, chen2023end}. \added{Alternatively, one can develop/compute rigorous bounds on the output of the ML model \cite{hertneck2018learning, paulson2020approximate} to guarantee constraint satisfaction.}

\subsubsection{ML for preprocessing}
Another key component of modern optimization solvers is preprocessing, a set of techniques used to reformulate the optimization problem and usually strengthen its relaxation \cite{achterberg2020presolve}. An example of a preprocessing procedure is bound tightening, where given a decision-making problem, the bounds of the variables are updated based on optimality and feasibility arguments. The former is known as Optimality Based Bound Tightening (OBBT), where given a decision-making problem, first the problem is convexified, and then the maximum and minimum value that a variable can take is found. This approach has been shown to lead to a reduction in solution time; however, it requires the solution of two optimization problems for each variable. ML has been used to determine the variables for which OBBT should be applied \cite{cengil2022learning}. This approach has been applied to the solution of optimal power flow problems where the ML model takes as input a vectorial representation of the parameters of the optimization problem and predicts the variables for which application of OBBT leads to the best bound. Finally, we note that a similar approach has been developed for the case of Feasibility Based Bound Tightening \deleted{(FBTT)} \added{(FBBT)} \cite{nannicini2011probing} \added{where the goal is to compute updated (tighter) bounds for all the variables while satisfying the constraints.}

\subsubsection{ML for branch and bound}
Branch and bound is the backbone of mixed-integer optimization solvers. In this approach, given a mixed-integer linear optimization problem of the form
\begin{equation} \label{full model MIP}
	\begin{aligned}
		      \minimize_{x,y} \ \ & c_{1}^{\top} x + c_{2}^{\top} y \\
			\subt \ \ & A_{1} x + A_{2} y \leq b\\
			& x \in \{0,1\}^{N_d}, y \in \R^{N_c},
		\end{aligned}   
\end{equation}
branch and bound starts by solving the continuous relaxation, i.e., setting $x\in [0,1]^{N_{d}}$ in Eq.~\ref{full model MIP}. The solution to this problem is usually fractional, i.e., the values of the $x$ variables are not integers. In this case, first, a variable $x_{i}$ is selected, and two new problems are created: one where $x_{i}$ is fixed to zero and one where $x_{i}$ is fixed to one (see Fig.~\ref{fig:bnb tree}). The procedure of selecting a variable to branch is known as variable selection. Once these two problems are generated, one has to select which one to solve (node $P_{1}$, $P_{3}$ or $P_4$ in Fig.~\ref{fig:bnb tree}); this is known as node selection. Overall, the variable and node selection strategies determine the computational efficiency of the branch and bound algorithm \cite{lodi2017learning}. Different variable selection rules have been proposed. A typical example is strong branching, where both branches corresponding to $x_{i}=0$ and $x_{i}=1$ are solved \added{for each and every variable $x_{i}$ with fractional value}, and the one providing the best bound is selected. Although this approach leads to smaller branch and bound trees, i.e., fewer nodes are explored, it is computationally expensive. Identifying the best variable and node strategy is an algorithm configuration problem. 
\begin{figure}[h]
    \centering
    \includegraphics{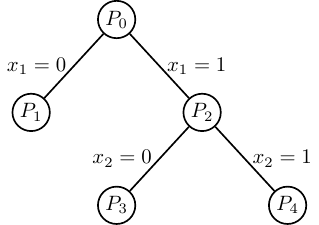}
    \caption{Branch and bound tree for a mixed integer linear optimization problem with two binary variables}
    \label{fig:bnb tree}
\end{figure}

Several ML-based approaches have recently been proposed to automate and reduce the computational effort related to making optimal decisions during the branch and bound solution process for mixed integer linear optimization problems \cite{lodi2017learning}. For variable selection, most approaches rely on the concept of imitation learning, where an ML model tries to copy the behavior of an expert, such as strong branching for the case of variable selection. This approach relies either on the vectorial representation of the problem \cite{khalil2016learning, alvarez2017machine} or the graph representation with features \cite{gasse2019exact}. An alternative approach is to exploit the sequential nature of variable selection and use reinforcement learning to find the variable to branch \cite{etheve2020reinforcement, scavuzzo2022learning, huang2022branch, parsonson2023reinforcement}. Finally, based on the relation between algorithm configuration and selection, the selection of a branching strategy has also been posed as an algorithm selection task for mixed integer linear problems \cite{di2016dash} and for spatial branching for polynomial optimization problems \cite{ghaddar2023learning}. 

Regarding node selection, two approaches have been proposed. In the first, node selection is posed as a Markov decision process and a policy is learned to determine which node to solve using imitation learning \cite{he2014learning, labassi2022learning}. The alternative is to pose node selection as a multiarm bandit problem, where given a set of options, one must select an option that will lead to the highest reward. In the context of node selection, the options correspond to the available nodes to explore, and the reward can be either the solution time or the size of the branch and bound tree to be explored \cite{sabharwal2012guiding}. 

\subsubsection{ML for cutting planes}
An important component of mixed-integer optimization algorithms/solvers is cutting planes \cite{dey2018theoretical}. These are usually linear inequalities that reduce the search space without affecting optimality. However, selecting which cutting plane to add is nontrivial since multiple types of cutting planes can be generated, and different numbers of cutting planes can be added during branch and bound. Mixed-integer optimization solvers create a pool of cuts and add them based on heuristics.

Similar to learning to branch, ML can be used to select which cuts to add. These approaches usually learn a model that approximates the outcome of an expert, i.e., a rule or heuristic, that identifies the best possible cut \cite{paulus2022learning, baltean2019scoring, marousi2022acceleration}. The addition of cutting planes can also be considered as a multistep process since cuts can be added to the root node as well as the other nodes that are explored during branch and bound. This has been considered in \cite{berthold2022learning}, where a regression model is used to determine whether using local cuts at a node of the branch and bound tree can lead to a reduction in solution time. The alternative is to rely on reinforcement learning to determine which cuts to add in each node of the branch and bound tree \cite{tang2020reinforcement, wang2023learning}.

\subsection{Configuring all the parameters of a solver simultaneously}

The ML approaches for algorithm configuration consider a specific aspect of the algorithm. One could consider all the parameters of a solver simultaneously. In this case, supervised, unsupervised, and reinforcement learning can be potentially used to identify the best configuration. Such approaches have been proposed for tuning mixed integer optimization solvers \cite{hutter2010automated,hutter2009paramils,bartz2004tuning, xu2011hydra, hutter2006performance, iommazzo2020learning}.

This approach can in principle exploit synergies between different parts of an algorithm or a solver. However, it leads to a significant increase in the complexity of the configuration and, subsequently, learning tasks. Furthermore, new architectures might be necessary to capture detailed information about the decision-making problem and the algorithm. For example, the graph representation with features and graph neural networks can guide the variable selection search during branch and bound. The algorithm, however, is usually represented as a vector, and each entry denotes the value of a hyperparameter. Therefore, new architectures and representations might be necessary to simultaneously capture information about the problem formulation and the algorithm configuration. \added{Finally, we note that these ML-based approaches usually can not provide guarantees regarding the performance of a solver or a configuration. This has motivated systematic analysis and design of numerical algorithms using data-driven \cite{balcan2024learning, sambharya2024data, dietrich2020koopman, doncevic2024recursively} and mathematical programming approaches \cite{drori2014performance, das2024branch, mitsos2018optimal}.}

\begin{table*}[h!]
\centering
\caption{Algorithm selection and configuration for different classes of optimization problems}
\setlength{\tabcolsep}{4pt} 
\renewcommand{\arraystretch}{1.2} 

\begin{tabular}{|p{1.75cm}|p{2cm}|p{3.5cm}|p{3.9cm}|p{3.75cm}|}
\hline
\multicolumn{2}{|c|}{\multirow{2}{*}{Task}} & \multicolumn{3}{c|}{Optimization problem class} \\ \cline{3-5}
\multicolumn{2}{|c|}{} & Continuous (Linear and Nonlinear) & Mixed Integer Linear & Mixed Integer Nonlinear \\ \hline
\multicolumn{2}{|c|}{Algorithm selection} & & & \\ \hline
\multirow{5}{*}{\parbox{3cm}{Algorithm\\ configuration}} 
& Initialization & \cite{vaupel2020accelerating, kumar2021industrial, baker2022emulating, park2023self, qian2024exploring} & 
\begin{tabular}[c]{@{}p{4cm}@{}}
\cite{masti2020learning, zhu2020fast, cauligi2022prism, russo2023learning, klauvco2019machine}\\
\cite{cauligi2021coco, misra2022learning, bertsimas2022online, triantafyllou2024deep}
\end{tabular} & \\ \cline{2-5}
& Preprocessing & \cite{cengil2022learning} & & \cite{nannicini2011probing} \\ \cline{2-5}
& \parbox{3cm}{Branching \\ priority} & \cite{ghaddar2023learning} & 
\cite{khalil2016learning, alvarez2017machine, gasse2019exact, etheve2020reinforcement, scavuzzo2022learning, huang2022branch, parsonson2023reinforcement} & \\ \cline{2-5}
& \parbox{3cm}{Node\\ Selection} & & \cite{he2014learning, labassi2022learning, sabharwal2012guiding} & \\ \cline{2-5}
& Cutting planes & & \cite{paulus2022learning, baltean2019scoring, marousi2022acceleration, berthold2022learning, tang2020reinforcement, wang2023learning} & \\ \cline{2-5}
& \parbox{3cm}{Multiple\\parameters} & \cite{chen2011random} & 
\cite{hutter2010automated, hutter2009paramils, bartz2004tuning, xu2011hydra, hutter2006performance, iommazzo2020learning} & 
\cite{liu2019tuning} \\ \hline
\end{tabular}
\end{table*}

\section{Learning to configure decomposition-based algorithms} \label{tune decomposition algorithms}
Decomposition-based optimization algorithms have been extensively used to solve large-scale decision-making problems. Unlike monolithic approaches where all the variables are considered simultaneously, decomposition-based algorithms decompose the variables (and constraints) into a number of subproblems that are solved repeatedly. Most decomposition-based algorithms can be classified either as distributed or hierarchical. The main difference lies in the sequence upon which the subproblems are solved. In distributed algorithms, all the subproblems are solved in parallel and are coordinated via dual information, whereas in hierarchical algorithms, the subproblems are solved sequentially and are coordinated either via dual information or cuts (for the case of cutting plane-based algorithms). In general, the solution of a decision-making problem with a decomposition-based algorithm has three steps: 1) problem decomposition, 2) selection of coordination scheme, and 3) configuration. These steps can be considered as hyperparameters of a decomposition-based algorithm; therefore, several algorithm configuration problems must be solved prior to the implementation of decomposition-based algorithms.

\subsection{Learning the structure of an optimization problem}
The decomposition of an optimization problem is the basis for the application of a decomposition-based optimization algorithm and can have a significant effect on the computational performance of the algorithm. Traditionally, a decomposition was obtained using intuition about the coupling (structure) among the variables and constraints. Although this approach has been applied extensively, identifying the underlying structure of a problem is time-consuming and may not even be possible using only intuition. 

Several automated structure detection methods have been proposed in the literature. These approaches rely on the graph representation of an optimization problem as represented in Section~\ref{sec: graph representation}. Given the graph of a decision-making problem, graph partitioning algorithms are used to decompose the graph, i.e., the decision-making problem, into a number of subproblems. Typical algorithms include hypergraph partitioning \cite{michelena1997hypergraph, wang2013computational, ferris1998partitioning, aykanat2004permuting, jalving2022graph} and community detection \cite{bergner2015automatic, khaniyev2018structure, tang2018optimal, allman2019decode, del2016automated, mitrai2020decomposition}. These graph partitioning methods usually make a-priory assumptions about the number of subproblems and the interaction patterns among them. To overcome these limitations, we have recently proposed the application of stochastic block modeling and Bayesian inference for estimating the structure of an optimization problem \cite{mitrai2022stochastic, mitrai2021iecr}. This approach assumes that the graph of an optimization problem is generated by a probabilistic model with parameters $b$ that capture information about the partition of the nodes into blocks \deleted{as well as} \added{and $\omega$ which captures} \deleted{the} interaction pattern between the blocks. \added{The parameter $b$ is a vector where the $i^{th}$ entry denotes the block membership of node $i$ in the partition of the graph. For the variable graph, this parameter denotes the block membership of each variable, whereas in the constraint graph, the block membership of each constraint.} Given the graph of a decision-making problem, the parameters $b$ are estimated or `learned' via Bayesian inference. The estimated structure can be used as the basis for the application of distributed and hierarchical decomposition-based algorithms. 

Finally, we note that regarding problem decomposition, the aforementioned approaches rely on the assumption that decomposing a decision-making problem based on the underlying structure leads to good computational performance. Although this has been shown to be a good assumption for a large class of problems \cite{tang2018optimal, basso2020random}, it is not guaranteed that a structure-based decomposition is the best possible one. An example is the case of the solution of two-stage stochastic optimization problems using Benders decomposition. Traditionally, the original problem is decomposed into a master problem, which considers the first stage decisions and a set of independent subproblems, each one representing a scenario. Recently it has been shown that adding some scenarios (or subproblems) to the master problem can lead to a reduction in the solution time \cite{crainic2021partial}. In general, finding the best possible decomposition is an open problem. 

\subsection{Learning to warm-start decomposition-based optimization algorithms}
Similar to the initialization of monolithic algorithms, the initialization of a decomposition-based algorithm can significantly affect its computational performance. However, predicting only the values of the variables is not enough since it does not account for the coordination aspect of a decomposition-based algorithm.

\subsubsection{Initialization of distributed algorithms} \label{subs: initialize distr algs}
For distributed-based algorithms, the coordination is achieved using Lagrangean multipliers. Therefore, an initialization requires an estimate of the values of the variables of the problem as well as the Lagrangean multipliers. This increases the complexity of the learning task compared to the initialization of monolithic solvers. This approach has been used to initialize the Lagrangean relaxation algorithm (a distributed decomposition-based algorithm) to solve network design and facility location problems \cite{demelas2023predicting}. This is achieved using an encoder-decoder architecture, where the input is the graph representation of the problem with features and the solution of the linear relaxation, and the output is an estimate of the multipliers. A similar approach has been developed for accelerating the Alternating Direction Methods of Multipliers (ADMM) using a recurrent neural network to predict the values of the Lagrangean multipliers and the complicating variables for the solution of optimal power flow problems \cite{biagioni2020learning}.

\subsubsection{Cutting plane-based hierarchical decomposition algorithms} \label{subs: initialize cut algs}

Initialization is more challenging for cutting plane-based decomposition algorithms. In these methods, a decision-making problem is usually decomposed into a master problem, which contains all the integer variables and potentially some continuous, and a subproblem, which considers only continuous variables. The solution of the subproblem depends on the values of the variables of the master problem, which are called complicating variables. The master problem and subproblem are solved sequentially and are coordinated via cutting planes, i.e., linear inequalities that inform the master problem about the effect of the complicating variables on the subproblem. Usually, in the first iteration, the master problem problem is solved without cuts; the cuts are added iteratively based on the solution of the master problem. Adding an initial set of cuts can lead to better bounds and, thus, convergence in fewer iterations. However, similar to the cutting plane methods for branch and bround, determining which cuts to add as a warm start for decomposition-based methods is nontrivial. First, the number of potential cuts can be very large, and selecting which ones to add is a complex combinatorial problem. The second issue is related to the validity of the cuts for different instances. For cases where the parameters of the subproblem do not change, the cuts can be evaluated only once and added to the master problem every time a new instance must be solved. However, if the parameters of the subproblem change, then the previously evaluated cuts are not valid. Thus, one has to evaluate them, i.e., solve the subproblem, before adding them to the master problem. 

In recent work, we have proposed several ML-based approaches to learn to initialize Benders decomposition by adding an initial set of cuts in the master problem for the solution of mixed integer model predictive control problems. For cases where the parameters of the subproblem do not change, and the complicating variables are continuous, we posed the cut selection problem as an algorithm configuration problem \cite{mitrai2024taking}. The number of cuts corresponds to the number of points used to discretize the domain of the complicating variables, and the performance function was the solution time, which was learned via active and supervised learning. 

For the case where the parameters of the subproblem do not change, and the complicating variables are both discrete and continuous, the cut selection process has two steps. First, an ML-based branch and check Benders decomposition algorithm is used to obtain an approximate integer feasible solution and a set of integer feasible solutions, which are explored during branch and check \cite{mitrai2024computationally}. The cuts related to these integer feasible solutions are added to the master problem, and then Benders decomposition is implemented to obtain the solution of the problem \cite{mitrai2024learning}. The integer feasible solutions guide the selection of the cuts to be added to the master problem. Finally, in the most generic case where the parameters of the subproblem change and the complicating variables are both continuous and discrete, a similar approach can be followed, where first, a set of integer feasible solutions is obtained by the ML-based branch and check. However, since the parameters of the subproblem change, the cuts related to these integer feasible solutions are first evaluated by solving the subproblem and then added to the master problem. 

\subsection{Learning to coordinate cutting plane-based decomposition algorithms}
Once a decomposition is decided and an initialization is selected, the next step is the implementation of the algorithm. As discussed in Sections~\ref{subs: initialize cut algs}, for cutting plane-based algorithms, the steps are 1) solve the master problem and obtain the values of the complicating variables, 2) solve the subproblem, and 3) incorporate \added{in the master problem information on the subproblems in the form of cuts}. \deleted{the information in the master problem for the subproblems, i.e., add cuts in the master problem.} These three steps are repeated until the algorithm converges. Selecting which cutting planes to add during the solution process is an algorithm configuration problem.

In certain classes of problems, multiple subproblems can exist, and in each iteration, multiple cuts can be generated and added to the master problem. Although this strategy seems reasonable at first since a cut contains information about the subproblem, it can also significantly increase the computational complexity of the master problem. This has led to the development of ML-based architectures to determine which cuts to generate and add to the master problem during the solution. Two approaches have been developed to achieve this. 

In the first, a classifier is used to predict whether a cut is valuable and should be added to the master problem. Different metrics are proposed to deem a cut valuable. The most commonly used one is the improvement in the bounds. This approach has been applied for the solution of two-stage stochastic optimization problems using Benders and generalized Benders decomposition \cite{jia2021benders, lee2020accelerating} as well as the solution of multistage stochastic optimization problems \cite{borozan2023machine}.  We note that a similar approach has been proposed for column generation where an ML model predicts if a column can lead to improvements in the bounds \cite{morabit2021machine}. 

The second approach exploits the iterative nature of decomposition -based algorithms and poses the cut selection problem as a reinforcement learning problem \cite{chi2022deep}. Specifically, the solution of a decision-making problem with a decomposition-based algorithm is modeled as a Markov Decision Process, and the goal is to train a reinforcement learning agent which given a candidate set of cuts (obtained from the solution of the master problem) selects the cuts that should be added such that the number of steps (iterations) required to solve the problem is minimized.

\begin{table*}[h!]
\centering
\caption{Algorithm selection and configuration for decomposition-based optimization algorithms}
\setlength{\tabcolsep}{4pt} 
\renewcommand{\arraystretch}{1.2} 

\begin{tabular}{|p{2.6cm}|p{4.5cm}|p{4cm}|p{4cm}|}
\hline
 \multirow{2}{*}{\textbf{Task}} & \multicolumn{3}{c|}{\textbf{Decomposition-based algorithm}} \\ \cline{2-4} 
 & \parbox{4.5cm}{Benders and Generalized \\ Benders Decomposition} 
 & \parbox{4cm}{Column Generation} 
 & \parbox{4cm}{Lagrangean Decomposition} \\ \hline

\parbox{3.5cm}{Structure detection} & 
\multicolumn{3}{p{12.5cm}|}{\parbox{12.5cm}{\cite{michelena1997hypergraph, wang2013computational, ferris1998partitioning, aykanat2004permuting, jalving2022graph, bergner2015automatic, khaniyev2018structure, tang2018optimal, allman2019decode, del2016automated, mitrai2020decomposition, mitrai2022stochastic, mitrai2021iecr, tang2018optimal, basso2020random}}} \\ \hline

Initialization & 
\cite{mitrai2024taking, mitrai2024learning} & 
& 
\cite{demelas2023predicting, biagioni2020learning} \\ \hline

\parbox{3.5cm}{Coordination via \\cutting planes} & 
\cite{jia2021benders, lee2020accelerating} & 
\cite{morabit2021machine, chi2022deep} & 
\\ \hline
\end{tabular}
\end{table*}

\section{Open problems and conclusions} \label{sec: conclusions}
In this section, we discuss open problems and new opportunities for applying ML to enhance the computational performance of algorithms for executing computational tasks in chemical engineering. 

\subsection{Application to general numerical tasks}
The concepts discussed in this paper, as well as the ML-based solution strategies, can be applied to generic computational tasks that arise in chemical engineering. Typical examples include steady-state and dynamic process simulation. In such cases, one must solve a system of equations using an iterative numerical algorithm that has hyperparameters. Hence, algorithm selection and configuration approaches can be used to select the best simulation algorithm and tune it for the specific computational task. Some examples include the tuning of the successive over-relaxation algorithm for the solution of linear systems of equations \cite{khodak2023learning}, selecting solvers for the solution of linear systems of equations \cite{demmel2005self, bhowmick2009towards, dufrechou2021machine} and for the solution of initial value problems \cite{kamel1993odexpert}. These results show that ML, in tandem with appropriate representations, might be able to accelerate process simulation, especially for large-scale and nonlinear systems, which are common in chemical engineering applications.

\subsection{Can ML generate new insights?} \label{subs: new insights}

All the aforementioned ML-based algorithm selection and configuration approaches answer the question of which algorithm to use and how to tune it. The next question is why is an algorithm (or configuration) able to solve a given problem instance efficiently? In other words, can ML generate new insights regarding the efficiency of a given algorithm for a class of decision-making problems? This question is relevant not only in the context of optimization algorithms but for the execution of numerical tasks in general \cite{kotthoff2016algorithm}. An approach to understanding the difficulty of solving a problem is to approximate the performance functions with interpretable models, such as decision trees, linear regression, and symbolic regression. However, these models usually have low accuracy, and more accurate models usually rely on deep learning (graph neural networks, feedforward neural networks, etc.), which is not inherently interpretable. This necessitates the utilization of explainable artificial intelligence tools for analyzing the outputs of deep learning models and potentially developing new interpretable deep learning architectures \cite{rudin2019stop, li2018deep}. Overall, explaining and understanding the computational performance of an algorithm for a given decision-making problem is an open research problem. 

\subsection{Data availability}
The data generation process is usually the most time 
consuming step in the development of an automated algorithm selection or configuration framework since a large number of decision -making problems must be solved, usually to optimality. Although parallel computing can be used to generate such datasets, this still requires significant computational resources. This computational cost can be potentially reduced using active, semi-, self-, and transfer learning approaches.

Active learning is a commonly used approach for cases where obtaining the labels of a data point is expensive \cite{settles2009active}. This approach has been used to learn to initialize Generalized Benders Decomposition for the solution of mixed-integer model predictive control problems \cite{mitrai2024initialize}. In this setting, a pool of data points is available, but only the features are known (e.g., some representation of the decision-making problems and the tuning) and obtaining the label requires the solution of the decision-making problem. The selection of the data point to be labeled is guided by the uncertainty of the prediction, i.e., we label the data point (combination of decision-making problem and tuning) for which the prediction of the solution time is the least certain. This approach still requires the labeling of data points.

Semi-supervised learning uses simultaneously labeled (usually few) and unlabeled data to train a ML model \cite{van2020survey}. An example is wrapper methods where a first model is trained using the labeled data (initial training set). The model is subsequently used to general pseudo-labels for the unlabeled data which are added to the training data set and the model is retrained. Self-supervised learning  uses the available unlabeled data to learn representations that can be useful for subsequent tasks such as classification and regression. Finally, transfer learning can be used to reduce the size of the training dataset by exploiting ML models trained for similar tasks \cite{weiss2016survey}, such as branching in mixed integer linear and mixed integer nonlinear optimization problems.

\subsection{Generative Artificial Intelligence}
All the ML-based methods discussed so far are based on predictive machine learning/ artificial intelligence techniques, namely supervised, unsupervised, and reinforcement learning. Recently generative artificial intelligence has made significant progress in developing AI-based systems capable of generating new content, such as video, image, and text. Given this remarkable progress, it is natural to wonder whether generative AI can be used to accelerate the solution of a decision-making problem.

The first application of generative AI is problem formulation from a natural language description of a decision-making problem. Preliminary results show that Large Language Models (LLMs) can successfully formulate an optimization problem when the number of parameters, variables, and constraints is small \cite{ramamonjison2022augmenting, ramamonjison2023nl4opt}. The natural language description has also been used to analyze infeasibility in a decision-making problem by making the LLM model interact with an optimization solver \cite{chen2023diagnosing}. LLMs have also been used to learn or discover new algorithms \cite{romera2024mathematical} by coupling an LLM with a genetic programming framework, where the LLM provides new candidate algorithms which are evaluated and subsequently mutated by the LLM. The last application is that of generating optimization instances. This is achieved using the graph representation, with node and edge features, of a decision-making problem and developing a model that generates new graphs, i.e., optimization problems \cite{geng2024deep}. 

Overall, generative AI can be conceptually used for problem formulation, explaining the solution of a computational task, discovering new algorithms, and reformulating a decision-making problem. However, the capability of current transformer-based deep learning architectures (both ones depending on natural language and graph-based) to perform these tasks is an open problem.

\section{Acknowledgments}
Financial support from NSF-CBET (award number 2313289) is gratefully acknowledged.

\bibliographystyle{elsarticle-num}
\bibliography{sample}

\end{document}